# An efficient method to calculate the anharmonicity free energy

Zhongqing Wu   and   Renata. M. Wentzcovitch

(Department of Chemical Engineering and Materials Science and Minnesota Supercomputing Institute, University of Minnesota, Minneapolis, MN 55455, USA)

The anharmonicity resulted from the intrinsic phonon interaction is neglected by quasiharmonic approximation. Although the intensive researches about anharmonicity have been done, up to now the free energy contributed by the anharmonicity is still difficult to calculate. Here we put forward a new method that can well include the anharmonicity. We introduce the implicit temperature dependence of effective frequency by volume modification. The quasiharmonic approximation becomes a special case in our method corresponding to non volume modification. Although our method is simple and only a constant need to determine, the anharmonicity is well included. Thermodynamic properties of MgO predicted with our method are excellent consistent with the experiment results at very wide temperature range. We also believe that our method will be helpful to reveal the characteristic of anharmonicity and intrinsic phonon interaction.

The harmonic approximation has been widely used to calculate the vibrational free energy. In this approximation there is no phonon interaction, the vibrational free energy can be expressed analytically by the phonon frequency. Quasi harmonic approximation (QHA) further considers the volume dependence of frequencies but ignored the temperature effect on frequencies. Base on the knowing of the vibrational spectrum obtained by experiment or calculated precisely with numerical methods such as density functional perturbation theory or frozen phonon method, QHA has been found to be able to produce thermodynamic properties well consistent with the experiment results especially at high pressure [1-5]. However, many investigations also discover that the QHA is inadequate in the region of low pressure and high temperature [6-13]. For example, at zero pressure, an obvious deviation of thermal expansivity of MgO [6] from experiment data begin to appears at about 400K. These results demonstrate that, at high temperature, the intrinsic phonon interaction (anharmonicity) neglected by QHA become prominent in many materials.

Although anharmonicity has been investigated intensely [3,7,8,14-17], the effectively method to include the anharmonicity in free energy is still absent. The intrinsic phonon interaction is very complex and change with temperature, which lead to it is not easy to include anharmonicity in free energy. The early approach [18,19], the anharmonic part of Helmhotltz free energy is written in the form of an expansion in terms of powers of T. The drawbacks of the method are that it introduces many volume-dependent adjustable parameters. There are not suitable manner to determine these parameters and the effect is also not very ideal. Even so, the method is still adopted to correct the QHA[10,20] up to now, which reflect partly the difficulty to deal with anhamonicity.

Here we put forward an improved method to QHA. The new method is very similar to quasi-harmonic approximation but can consider well the anharmonic effect.

In the quasiharmonic approximation(QHA), the Helmholtz free energy is given by

$$F_0(V,T) = U_0(V) + \frac{1}{2}\sum_{q,j} h\omega_j(q,V) \\ + k_B T \sum_{q,j} \ln\{1 - \exp[-h\omega_j(q,V)/k_B T]\} \quad (1)$$

where the first term is internal contribution, the secondly term and third term is the zero point and vibration contribution respectively. The phonon frequency is only determined by volume and independent of the temperature in quasiharmonic approximation. The intrinsic phonon interaction, which changes with temperature, has been neglected. Here we think, as in quasi particle theory, the interaction phonon can be renormalized into non-interaction phonon with an effective frequency. We assume

$$\omega_{eff}(V,T) = \omega(V',0) \quad (2)$$

Where,

$$V' = V(P,T)\{1 - c*[V(P,T) - V(P,0)]/V(P,0)\} \quad (3)$$

$c$ is constant, the second term is a temperature-dependent volume modification. Through this volume modification, we introduce the temperature-dependence of the effective phonon frequency. Then the Helmholtz free energy can be rewritten as

$$\begin{aligned}F_c(V,T) &= U_0(V) + \frac{1}{2}\sum_{q,j} h\omega_{eff\ j}(q,V,T) + k_B T \sum_{q,j} \ln\{1 - \exp[-h\omega_{eff\ j}(q,V,T)/k_B T]\} \\ &= U_0(V) + \frac{1}{2}\sum_{q,j} h\omega_j(q,V') + k_B T \sum_{q,j} \ln\{1 - \exp[-h\omega_j(q,V')/k_B T]\} \\ &= U_0(V) + F_0(V',T) - U_0(V')\end{aligned} \quad (4)$$

After this generalization, QHA become a special case according to c=0. For calculation, firstly, we apply QHA to get $F_0(V,T)$ and $V_0(P,T)$ then we can obtain V' according to equation (3) and further get $F_c(V,T)$ from equation (4).

In order to demonstrate the effectivity of the method, we investigate the thermodynamic of MgO. MgO has a salt structure and is one of the end members of magnesiowüstite that is expected to the important constituent of Earth lower mantle. The detail of calculation here is very similar to karki et al [6]. Computation are performed using the density functional theory with the local density approximation [21] and the PWscf code[22]. Mg pseudopotential is generated by the method of von Barth and Car [23]. O pseudopotential is generated by the methods of Troullier and Martins [24]. The plane wave energy cutoff is 70 Ry. Brillouin zone sampling for electronic states was carried out on 10 k-points. Phonon frequencies were calculated using density functional perturbation theory [25,26]. Then thermodynamic properties are determined by the method above mentioned.

Although the method is very simple, it can well include the anharmonic effect. This is very obvious in thermal expansivity of MgO indicated Fig 1. Comparing the result for different c, we can find that difference between $c \neq 0$ and $c=0$(QHA) can be ignored at low temperature and become prominent at high temperature. The difference also becomes less and less pronounced at high pressure. All characteristic are very similar to anharmonic effect. In fact, at P=100GPa the difference can almost be overlooked. This is well consistent with the Inbar and Cohen [8] result in which they don't find significant difference between molecular dynamics (including anharmonic effect) and QHA results at such pressure. The inclusion of anharmonicity tends to give rather slow and linear high temperature dependence of thermal expansivity. By increasing the parameter from c=0 to c=0.1, our result become more and more consistent with the experiment. The result for c=0.1 is in excellent consistent with the experiment results. The difference between c=0.1 and experimental results is less than 1% even at T=1700K, which is in the range of experimental error. The result is also well consistent with molecular dynamics simulations based on variational induced breathing (VIB) model, which includes anharmonicity. At zero pressure, the anharmonic effect is very large at high temperature. The difference between c=0 (QHA) and c=0.1 results is about 4% at T= 800K whereas it become 9% at T=1500K. The anharmonic effect are still obvious even at P=30GPa, where the difference between c=0.1 and QHA is about 5% at T=2000K.

QHA also overestimates heat capacity at constant pressure ($C_p$) at high temperature. The $C_p$ of P=10GPa instead of P=0GPa agree well with experimental results at high temperature and zero pressure [6]. The overestimation can be reduced by our method, whose result is shown in Fig. 2. With increasing c from 0 to 0.1, the consistent between the calculation and experiment become better. The difference between c=0.1 and experimental results is in the range of the experimental errors. Similar to the result of thermal expansivity, we can found from $C_p$ that the anharmonic effect begin to be obvious above 400K at zero pressure.

The Grüneisen parameter is a very important parameter to calculating the thermal pressure. It is very useful in reducing iso-temperature equation of state from shock wave experiments. It is also a useful indicator for the importance of anharmonic effect. In the experimental results, the Grüneisen parameter is almost temperature independent. However γ of QHA increase rapidly in high temperature. The strong temperature dependence of Grüneisen parameter is attributed to the neglect of anharmonicity in QHA. This is very obvious in the temperature and pressure dependent of Grüneisen parameter with different c shown in Fig.3. With increasing the c, the temperature – dependent of Grüneisen parameter become weak. The result of c=0.1 is almost temperature independent. Furthermore the values of Grüneisen parameter are more consistent with the experiment. The Gruneisen parameter for c=0.1 is 1.534 at T=500K, which is almost the same with the experimental results 1.53. The difference between Grüneisen parameter for c= 0.1 and experimental result is less than 1% up to T>1000K. The result of molecular dynamics simulation by Inbar *et.al.* [8] (including the anharmonic effect ) also shown in Fig.3 for comparison. Our result for c=0.1 has the similar temperature independent relation to molecular dynamic. But the value of molecular dynamics simulation is far below than the experiment result.

Now we know that the improvement of our method to QHA is very obvious. The results predicted with our method can be excellent consistent with the experiment results at very wide temperature range. The parameter c can be determined by comparing to the experiment result. The value for MgO from different experiment such as thermal expansivity, Grüneisen parameter and heat capacity are almost the same. c=0.1. Therefore, our method systemically improves the results of QHA. Furthermore, after the improvement, our result is more consistent with the experiment than molecular dynamic simulation [8] including the anharmonicity. This is particularly apparent in Grüneisen parameter. Both our result and molecular dynamic simulation show the weak temperature-dependent of Grüneisen parameter. But molecular dynamic simulation obviously underestimates Grüneisen parameter.

We think, as in quasi particle theory, the interaction phonon can be renormalized into non-interaction phonon with an effective frequency. Due to the complicacy of phonon interaction, it is not easy to get the explicit temperature dependence of the effective phonon frequency. However, the implicit temperature dependence of effective frequency, which introduced by volume modification shown in equation (3), has been found to be able to include the anharmonicity very well. This can be understood. The anharmonicity is prominent at high temperature and become less and less pronounced with increasing pressure. This characteristic is embodied by volume modification in equation (3): The higher temperature is, the larger the volume modification is. The lower pressure is, the larger the volume modification is. It is obvious that there are many choices for volume modification meet above characteristic. Therefore it should be possible to improve further by adjust the volume modification in Equation (3) if someone needs a better high temperature result.

In conclusion, we put forward to a method which can effectively include anharmonicity. QHA become a special case in our method. The method is simple. Only a constant need be determined by the experiment. However, the improvement of our method to QHA is very obvious. Thermodynamic properties predicted with our method are excellent consistent with the experiment results at very wide temperature range. We also believe that our method should be also very helpful to deepen the knowledge of the nature of anhramonicity.

[1] T. Tsuchiya, J. Tsuchiya, K. Umemoto, and R. M. Wentzcovitch, Earth Planet. Sci. Lett. **224**, 241(2004).

[2] B. B. Karki, R. M. Wentzcovitch, S. de Gironcoli, and S. Baroni, Science **286** , 1705(1999).

[3] A. R. Oganov, J. P. Brodholt, and G. D. Price, Phys. Earth Planet. Inter. **122**, 277(2000),

[4] A. Chopelas, Phys.Earth Planet. Inter. **98,** 3(1996).

[5] O. L. Anderson and K. Masuda, Phys.Earth Planet. Inter. **85**, 227(1994).

[6] B. B. Karki, R. M. Wentzcovitch, S. De Gironcoli, and S. Baroni, Phys. Rev. B **61**, 8793(1999).

[7] M. Matsui, G. D. Price and A. Patel, Geophys. Rev. Lett **21**, 1659(1994).

[8] I. Inbar and R. E. Cohen, Geophys. Rev. Lett **22**, 1533(1995).

[9] B. Bertheville, H Bill, and H. Hagemann, J. Phys.: Condens. Matter **10**, 2155(1998).

[10] G. Kern, G. Kresse, and J. Hafner, Phys. Rev. B **59**, 8551(1999).

[11] A. R. Oganov, and S. Ono, NatureT **430**, 445(2004).

[12] S. M. Foiles, Phys. Rev. B **49**, 14930(1994)

[13] F. D. Stacey and D. G. Isaak, J. Geophys. Res. **108**, 2440(2003)

[14] J. P.K. Doye and D. J. Wales, J. Chem. Phys. **102**, 9659(1995)

[15] A. Miari et al., J. Chem. Phys. **112**, 248(2000)

[16] T. Lang, M. Motzkus, H. M. Frey, and P. Beaud J. Chem. Phys. **115**, 5418(2001)

[17] V. L. Gurevich, D. A. Parshin, and H. R. Schober, Phys. Rev. B **67**, 094203 (2003)

[18] L. D. Landau and E. M. Lifshitz, Course of Theoretical Physics, Statistical Physics, Part I (Butterworth & Heinemann, Oxford, 1980), Vol 5.

[19] D. C. Wallace, Thermodynamics of Crystal (Wiley, New York, 1972)

[20] A. R. Oganov and P. I. Dorogokupets, Phys. Rev. B **67**, 224110(2003).

[21] J. P. Perdew and A. Zunger, Phys. Rev. B **23**, 5048(1981).

[22] http://www.pwscf.org.

[23] U. von Barth and R. Car unpublished..

[24] N. Troullier, and J.L. Martins, Phys. Rev. B **43**, 1993(1991).

[25] S. Baroni, P. Giannozzi, and A. Testa,, Phys. Rev. Lett. **58**, 1861(1987).

[26] P. Giannozzi, S. de Gironcoli, P. Pavone, and S. Baroni, Phys. Rev. B **43** , 7231(1991).

[27] Y. S. Touloukian, R. K. Kirdby, R. E. Taylor, and T. Y. R. Lee, Thermophysical Properties of Matter ~Plenum, New York, 1977, Vol. 13

[28] D. G. Isaak, O. L. Anderson, and T. Goto, Phys. Chem. Miner. **16**, 704 (1989).

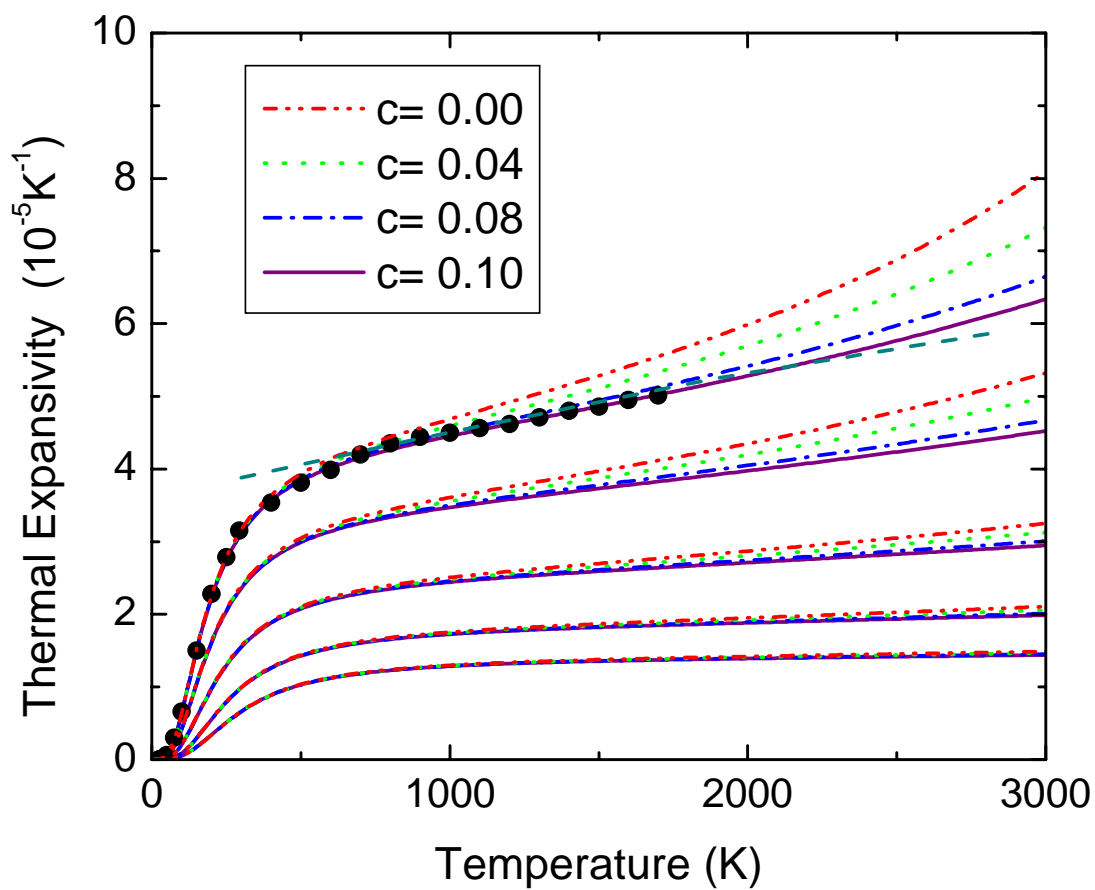

Fig. 1. Pressure and temperature dependence of thermal expansivity. The pressure is 0, 10, 30, 60, 100 GPa from top to bottom respectively. Experimental data at ambient pressure are denoted by circles[27] and Dashed lines from molecular dynamic simulation [8]. Parameter $c$ from Eq(3)

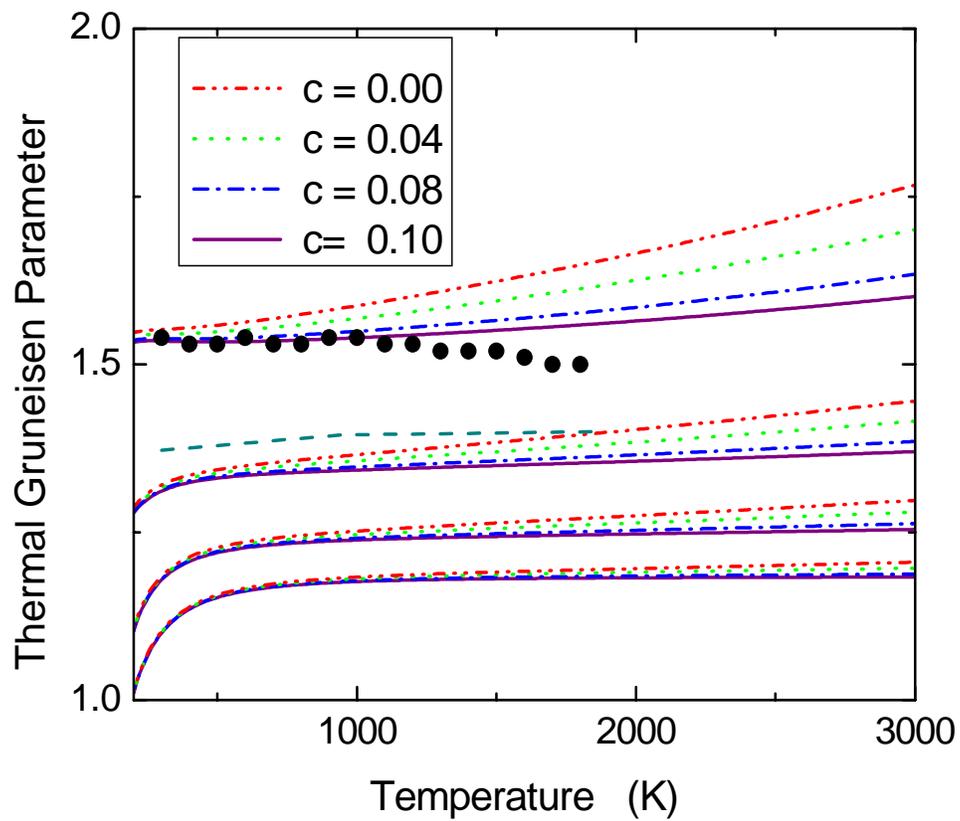

Fig. 2  Pressure and temperature dependence of thermal Grüneisen parameter. Pressure is 0, 30, 60, 100 GPa from top to bottom respectively. Experimental data at ambient pressure are denoted by circles [28]. Dashed lines from molecular simulation. Parameter $c$ from Eq(3)

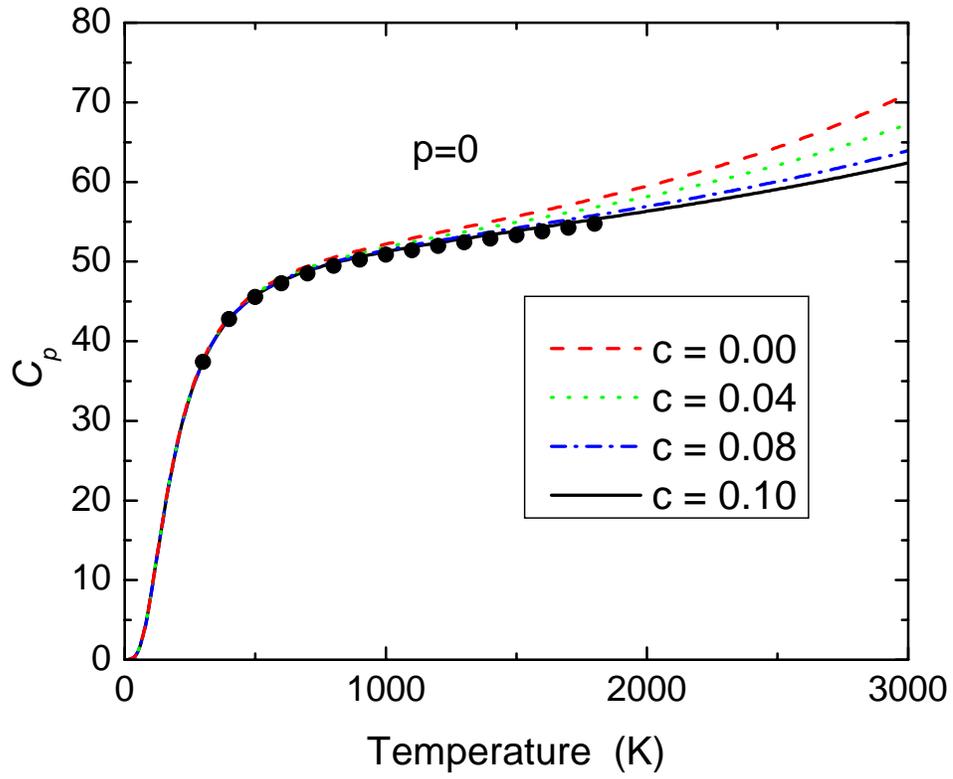

Fig. 3. Temperature dependence of isobar heat capacity at ambient pressure. Experimental data at ambient pressure are denoted by symbols[28]. Parameter c from Eq (3)